\documentclass[conference]{IEEEtran}
\IEEEoverridecommandlockouts

\usepackage{cite}
\usepackage{amsmath,amssymb,amsfonts}
\usepackage{algorithmic}
\usepackage{textcomp}
\usepackage{xcolor}
\def\BibTeX{{\rm B\kern-.05em{\sc i\kern-.025em b}\kern-.08em
    T\kern-.1667em\lower.7ex\hbox{E}\kern-.125emX}}
\usepackage{graphicx,url}
\usepackage[utf8]{inputenc}
\usepackage{framed}
\usepackage{booktabs}
\usepackage{helvet}
\usepackage[square,numbers,sort]{natbib}
\usepackage{epstopdf}
\usepackage{rotating}

\title{A Preliminary Search for Evidence on Government Software Engineering Practices: Results from Three Rapid Reviews}

\author{\IEEEauthorblockN{Sebastián Pizard}
\IEEEauthorblockA{\textit{Facultad de Ingeniería} \\
\textit{Universidad de la República}\\
Montevideo, Uruguay \\
spizard@fing.edu.uy}
\and
\IEEEauthorblockN{Matías Porro}
\IEEEauthorblockA{\textit{Facultad de Ingeniería} \\
\textit{Universidad de la República}\\
Montevideo, Uruguay \\
matias.porro@gmail.com}
\and
\IEEEauthorblockN{Andrea Muñoz}
\IEEEauthorblockA{\textit{Facultad de Ingeniería} \\
\textit{Universidad de la República}\\
Montevideo, Uruguay \\
amuniozv@gmail.com}
\and
\IEEEauthorblockN{Andrea Delgado}
\IEEEauthorblockA{\textit{Facultad de Ingeniería} \\
\textit{Universidad de la República}\\
Montevideo, Uruguay \\
adelgado@fing.edu.uy}
}

\begin{document} 

\maketitle

\begin{abstract}
Government agencies are major software developers and drivers of digital transformation, yet empirical evidence on their software engineering (SE) practices remains largely unexplored. This paper reports three rapid reviews of 2024 peer reviewed publications to assess the availability of evidence on government SE practices. We examined a subset of top tier international SE venues, regional South American SE conferences, and a curated newsletter on public sector digitalization. Across 984 screened papers, we identified only four studies reporting on SE practices conducted by, for, or in collaboration with government bodies, most published in regional venues as experience reports or case studies. Our findings are consistent with practitioners' perceptions that academic evidence on government SE practices has low visibility in mainstream SE venues. We discuss implications including the need for dedicated publication venues, incentives for government academia collaboration, and improved methods for evaluating grey literature, currently a key information source for government agencies. This work provides a preliminary study for understanding the evidence landscape and identifies directions for future research to better support evidence based software engineering in the public sector.
\end{abstract}


\section{Introduction}\label{sec:intro}

In medicine, evidence-based practice (EBP\footnote{In this paper, we use EBP as a generic term when our comments concern the use of evidence in any applied discipline. We use EBSE when our comments concern the use of evidence only in the context of software engineering (SE), computer science (CS), and information technology (IT).}) has helped practitioners meet their information needs while maintaining research rigor \cite{goues2018}. Evidence-based software engineering (EBSE), its adaptation to our discipline, aims to improve decisions about software development and maintenance by combining the best available research evidence with professional expertise and human values. 

Systematic reviews (SRs\footnote{We use the term SR to refer to any form of systematic review, this includes mapping studies (MS), rapid reviews (RR) and tertiary studies (TS).}) are central to EBSE because they offer a transparent, methodical way to locate and combine the available research evidence  on a given topic or research question. SRs aggregate evidence from articles reporting individual empirical studies (called primary studies); the SRs themselves are secondary studies. Since EBSE's introduction in 2004 \cite{kitchenham2004}, SRs have become widely used in software engineering for synthesizing evidence across many topics. For instance, Kamei et al.identified 446 SRs published in leading SE journals and conferences before 2019 \citep{kamei2021}.

More than twenty years after EBSE was introduced, its adoption outside academia remains limited. Adopting EBSE in industry or government means using scientific evidence, particularly evidence synthesized in SRs, to inform decision making. Several factors contribute to this limited adoption. First, weak connections between academia and industry hinder knowledge transfer. Second, many SRs fail to address practitioners' questions, lack actionable recommendations, and serve primarily academic purposes \cite{hassler2014, cartaxo2016}. Third, some topics lack sufficient primary studies to support evidence synthesis. 



Unterkalmsteiner and Gorschek demonstrate that government agencies (GAs) can benefit from SE technologies for requirements specification, requirements management, and project management \cite{unterkalmsteiner2018}. However, they did not address how to select appropriate technologies, a question EBSE could help answer. GAs could serve dual roles: as users of EBSE in their own work and as advocates for EBSE by requiring empirical evidence to support their recommendations. 

A recent study on EBSE adoption in government agencies \citep{pizard2023} suggests that some GAs related to SE may be interested in adopting evidence-based practice to strengthen or complement their research and policy-making processes. However, it also points to a potentially important barrier: there may be limited scientific evidence about GAs’ SE practices (i.e., empirical findings on how government agencies develop, procure, evolve, and manage software), which is precisely the evidence that would be most useful for GAs' interests.

Motivated to better understand this potential barrier, our goal is to assess the state of evidence on GAs' SE practices. As preliminary first step toward this objective, we carried out three Rapid Reviews (RRs) —a lightweight form of SR designed for resource-constrained settings \citep{hamel2021}—to identify peer-reviewed publications on government SE practices published in 2024. The first targeted top-tier SE journals and conferences, the second focused on regional SE venues in South America, and the third screened a newsletter that curates academic papers on government digitalization. Across nearly one thousand papers, we found only four that reported on GAs' SE practices. In this paper, we report the RRs conducted as well as a discussion and reflection on the results.

This paper makes the following contributions: (i) empirical evidence on the visibility of government SE practices in contemporary SE venues, showing that only four of 984 screened 2024 publications report such practices; (ii) a pragmatic example of how a set of complementary rapid reviews can be combined to address a practical, resource-constrained question, providing an approach that other researchers and practitioners can reuse; and (iii) a set of implications and future directions for better supporting EBSE in the public sector, including an initial exploration of how to appraise the grey literature that government agencies actually rely on.

This paper is structured as follows. Section \ref{sec:related} provides background on the relationship between software engineering and government agencies, and reviews related work on the use of evidence about software engineering practices in governmental settings. Section \ref{sec:method} states our research objectives and describes the research strategy used to address them. Section \ref{sec:results} presents and discusses the results, followed by Section \ref{sec:threats}, which outlines threats to validity of our study. Finally, Section \ref{sec:final} presents final remarks.

\section{Background \& Related Research} \label{sec:related}

Throughout the history of computing, government agencies have been central drivers of software engineering. Early large-scale, mission-critical systems in defense, aerospace, taxation, and census administration (such as the Semi-Automatic Ground Environment, SAGE) exposed recurring problems of cost overruns, unreliability, and unmanaged complexity that contributed to the \emph{``software crisis''} and to the formalization of the term \emph{``software engineering''} at the North Atlantic Treaty Organization (NATO) conferences of 1968--1969 \cite{campbell_kelly_2003, ceruzzi_1998, nato_software_1968}. Agencies such as the National Aeronautics and Space Administration (NASA) and the U.S. Department of Defense (DoD) later sponsored advances in requirements engineering and safety-critical development, and the Software Engineering Institute (SEI), created in 1984, institutionalized maturity models and evaluation methods that governments adopted to assess software suppliers \cite{sei_cmmi_2010}. 

From the late 1990s onward, the relationship expanded to digital government and citizen-facing services: international bodies such as the United Nations (UN), the Organisation for Economic Co-operation and Development (OECD), and the World Bank promoted e-government and digital-transformation frameworks, and national units such as the United Kingdom (UK) Government Digital Service and the U.S. Digital Service introduced agile methods, DevOps, open-source development, and user-centered design into the public sector \cite{un_egov_2022, oecd_digigov_2014, oecd_ai_2019}. In this contemporary phase, software engineering both enables and is shaped by government action: public policies define requirements for transparency, security, equity, and resilience, while engineering practices make it possible to build and maintain the large-scale digital systems that underpin modern governance \cite{sei_cmmi_2010, oecd_ai_2019}.

However, despite the strong influence of government action and public policy on SE, it is not easy to identify recognized empirical studies on SE practices from government agencies. This contrasts with the existence of widely recognized empirical work focused on SE industry practices (see, for example, \cite{sadowski2018, devanbu2016}). Empirical evidence about SE practices outside academic contexts is very important. It provides inputs and validation for research and education in the discipline. It also offers reference points and guidance for companies or public agencies that resemble the organizations that produced the evidence. In addition, it supports the adoption of EBSE by non-academic stakeholders because it connects them with evidence that comes from contexts similar to their own.

To the best of our knowledge, a recent study reports the first empirical investigation of EBSE adoption within GAs, conducted with a Uruguayan government agency \cite{pizard2023}. That agency, Uruguay's Agency for Electronic Government and Information Society and Knowledge (AGESIC)\footnote{https://www.gub.uy/agencia-gobierno-electronico-sociedad-informacion-conocimiento/}, is an executive body under the Presidency of Uruguay that leads the country's digital government strategy. Created in 2006, it aims to improve citizen services through information and communication technologies and to foster an inclusive, skills-oriented information society. AGESIC drives initiatives in digital service delivery, open government, data-driven decision making, interoperability, and digital security; it also implements national IT programs, sets technical standards, and plays a policy-making and regulatory role in SE by defining software requirements and procurement conditions, promoting software quality, and issuing recommendations on technology risks. As a result, Uruguay has led the UN e-Government index (EGDI) in Latin America for more than ten consecutive years, climbing in 2024 to second place in all of the Americas\footnote{https://desapublications.un.org/publications/un-e-government-survey-2024}, only behind the U.S.

The study examines stakeholders' perspectives and whether EBSE knowledge affects professional practice. Through a two-stage field study with AGESIC, the authors combined an EBSE awareness session and a focus group with a follow-up session 16 months later. Participants viewed EBSE as promising for complex policy and software-related decisions, but also identified several barriers to adoption, including limited institutional support, skill and motivation gaps, the cost of conducting systematic reviews, and a lack of evidence on emerging topics. Among these, one barrier is especially relevant to the present work: the scarcity of evidence about government software practices---precisely the kind of evidence that would be most useful to GAs. 

Digitalization agencies with roles similar to AGESIC exist in many countries, and other entities with related responsibilities, such as IT divisions within state-owned enterprises, face comparable challenges. One of their key tasks is to formalize and facilitate SE practices across government agencies. These agencies are interested in sharing experiences and generating cumulative knowledge. To this end, for example, the UK has a cross-government SE community \cite{ukgov2025}. Such agencies and communities would greatly benefit from empirical research on SE practices in government, including both SRs and primary studies. However, it remains unclear whether such evidence exists in systematic form or where it can be accessed.

\section{Goal and Methods} \label{sec:method}

Our goal was to assess the state of evidence on GAs' SE practices. Throughout this study, we operationalize \emph{government SE practices} as software engineering activities (e.g., requirements, architecture, process, testing, maintenance) carried out by, for, or in collaboration with government bodies (including ministries, public agencies, regulatory bodies, and state-owned enterprises) regardless of whether the work explicitly invokes EBSE. Specifically, we sought to examine the perception reported by practitioners in previous research \cite{pizard2023}, namely that little evidence exists on GAs' SE practices. As a preliminary step, we aimed to gauge the volume of this work relative to the broader SE literature and, where possible, to identify publication venues with a higher concentration of such studies. We posed the following research question:

\begin{framed}
   \emph{RQ: What evidence exists on governments' software-engineering practices?}
\end{framed}

We adopted a sampling approach focused on publications from 2024. Rather than attempting an exhaustive historical assessment, we aimed to obtain a snapshot of the most recent evidence available at the time of the study. This kept the effort compatible with the rapid-review methodology while allowing us to estimate how visible government software-engineering practices are in contemporary publication venues. Our object of study is the \emph{availability of evidence} on government SE practices, not the extent to which government agencies themselves adopt EBSE (the latter being the focus of prior work \cite{pizard2023}); accordingly, our counts reflect how visible such evidence is in the selected venues during 2024, not an estimate of the total volume of relevant work published worldwide. This study was primarily intended to inform our own decisions about related, ongoing research---in particular, our collaborations with government agencies and the design of a future, broader investigation. Nevertheless, we believe the findings may also be of interest to other researchers and practitioners concerned with the availability of evidence on government SE practices. 

Specifically, we conducted three rapid reviews (hereafter RR1, RR2, and RR3) to identify 2024 publications on government SE practices in:
\begin{enumerate}
    \item[i.] Four top-tier SE journals and conferences (RR1),
    \item [ii.] Two regional SE venues (RR2), and,
    \item [iii.] Issues of EDOS\footnote{Effective Digitalisation of the Public Sector (EDOS) is a weekly newsletter (published in English and Norwegian) that curates and highlights research and insights related to public-sector digitalisation, aiming to make evidence more accessible and useful for practitioners and decision-makers. \url{https://enedos.substack.com/}}, a Norwegian newsletter covering government digitalization (RR3).
\end{enumerate}

\subsection{Rationale for using Rapid Reviews} 

A rapid review is a type of knowledge synthesis that speeds up the traditional systematic review process by streamlining or omitting certain methodological steps to efficiently provide evidence for stakeholders while conserving resources~\citep{hamel2021}. These stakeholders, or knowledge users, include practitioners, policymakers, educators, public partners, and industry leaders; involving them helps ensure relevance and uptake \citep{garritty2024}. A key stakeholder is the requester, who defines the information needs and often supports validation and dissemination, and researchers may also play this role when an RR informs their future research or industry collaboration \cite{pizard2025}. Because there is no single RR method, approaches are tailored to needs and constraints, with success relying on early, continuous engagement with the requester and common streamlining tactics such as limiting searches, using single-reviewer processes with verification, focusing risk-of-bias on key outcomes, and favoring descriptive over quantitative synthesis \citep{king2022, garritty2020}. For all these characteristics, RRs seem a suitable method for exploring our problem, as they allow us to conduct several complementary preliminary investigations. 

\subsection{Rationale for the Selected Sources \& Period} 

The selected sources were not intended to represent the entire body of literature on government digitalization. Instead, they were chosen to cover three complementary perspectives: (i) leading empirical software-engineering venues, where rigorous SE evidence is most likely to be reported; (ii) regional venues, which increase geographical diversity and may capture studies not published internationally; and (iii) a curated source focused on public-sector digitalization, providing exposure to publications that may not typically appear in SE venues. Together, these sources were intended to provide an initial exploratory view of the availability of evidence on government software-engineering practices. The selection of venues was purposive and informed by the authors' experience rather than by a systematic database search; this is an accepted streamlining tactic in exploratory rapid reviews, at the cost of generalizability.

It is important to note that RR1 covers top-tier international venues (IST, ESEM, EMSE, ICSE), whereas RR2 covers regional venues (CLEI, CIbSE, SBES). Other prominent SE venues, e.g., IEEE Transactions on Software Engineering (TSE), Journal of Systems and Software (JSS), IEEE Software, EASE, and PROFES, were not included in this preliminary sample. We prioritized a small set of venues with a high prevalence of empirical studies to keep the effort compatible with the rapid-review methodology; the remaining venues are candidates for the planned extended search.

In RR3, EDOS was selected because it is a curated, scholarly newsletter explicitly maintained by academics and funded by a digitalization ministry, offering a structured and regularly updated entry point to public-sector digitalization research. We do not claim it is the only such source; broadening to additional newsletters/curated outlets is left for future work.

We intentionally excluded dedicated digital-government and public-administration venues (e.g., Government Information Quarterly (GIQ); Digital Government: Research and Practice (DGOV); the International Conference on Theory and Practice of Electronic Governance (ICEGOV); the International Conference on Electronic Government and the Information Systems Perspective (EGOVIS); and the International Journal of Electronic Government Research) from this first sample to keep the focus on SE-native venues, where SE practices are reported with engineering detail. These venues are a priority for the planned extended search.

We deliberately limited the scope to a single recent year (2024) as a snapshot consistent with the rapid-review methodology, prioritizing breadth of venue coverage over historical depth. We do not claim this year is representative of the entire evidence base; rather, it offers a current, high-visibility cross-section. Extending the window (e.g., 2020--2026) is part of our planned future work.

\subsection{Rapid Reviews Details} 

The searches were conducted manually, covering all publications from these sources and the newsletter during 2024. This enabled a preliminary, exploratory assessment aligned with our purpose. By manually screening all publications from the selected sources, we aimed to achieve comprehensive coverage of those venues within the selected time frame. This approach allows us to determine what proportion of publications in these sources actually report on GAs' SE practices, serving as an initial sample that provides insight into the current landscape.

Study selection and data extraction were performed by the second and third authors, under the supervision of the first author and with a final review by the fourth author. Both the second and third authors had prior training in evidence-based software engineering and secondary-study methods: the second author had taken an EBSE and SRs course \cite{pizard2021}, in which he participated in conducting a systematic review, and the third author was trained in EBSE and SRs before this research. Nonetheless, these were the first rapid reviews conducted by either of them (see Section \ref{sec:threats}). The first author, who has extensive experience in secondary studies, supervised the process throughout, reviewed decisions, and validated a sample of the data; the fourth author independently reviewed the three RRs and helped discuss their results. Responsibilities were distributed per review: in RR1, the second author performed the study selection and the first author validated the screening of the first 20 records; in RR2, the third author performed the selection and the first author revised the outcomes; and in RR3, the second author performed the selection and the first author independently classified a random sample to validate the results. Whenever uncertainties arose regarding study eligibility or classification, decisions were discussed among the researchers until consensus was reached.

Tables \ref{tab:1rr}, \ref{tab:2rr}, and \ref{tab:3rr} present the key characteristics of the RRs we conducted\footnote{Supplementary information in https://zenodo.org/records/18099188}. To comply with the proposed key information to report in RRs \cite{cartaxo2020}, the following items, which apply to all three RRs, should be added:
\begin{itemize}
    \item \emph{Problem:} Identify and report government software engineering practices, that is, SE studies conducted by, for, or in collaboration with governmental bodies.
    \item \emph{Protocol:} In all three cases, the protocol was drafted during the first week, jointly by the reviewer (the second or the third author) and the first author.
    \item \emph{Stakeholders’ roles:} The requester was the first author, who will use the RR results to improve his collaborations with AGESIC and Uruguayan government agencies.
    \item \emph{Evidence Appraisal:} No quality assessment was planned. 
    \item \emph{Report and diffusion:} We did not preplan dissemination effort because the evidence was intended for internal use.
\end{itemize}

\begin{table}[htbp!]
	\caption{Key features of the RR on government SE practices in top-tier venues (RR1).}
	\label{tab:1rr}
\begin{tabular}{p{1.1cm}p{6.7cm}}
\toprule
Research Question    & What evidence on government practices was published in top-tier SE venues during 2024?   \\ \hline
Time Frame           & Approximately 40 hours.   \\ \hline
Search Strategy      & Manual search of papers published in 2024 in the following venues: Information and Software Technology (IST) Journal, International Symposium on Empirical Software Engineering and Measurement (ESEM), Empirical Software Engineering (EMSE) Journal, and the International Conference on Software Engineering (ICSE). These venues were chosen after ranking top-tier outlets by the prevalence of empirical studies.  \\ \hline
Selection Procedure  & We considered studies conducted by, for, or in collaboration with governments (mentions of government, specific countries, public agencies, regulatory bodies, etc.), written in English, with full text available. The second author performed the study selection, and the first author reviewed a sample of the first 20 records to validate the screening.     \\ \hline
Extraction Procedure & The predefined extraction and synthesis plan included collecting publication data (title, authors, etc.), countries of author affiliations, and classifying items by topics of interest, for example SE subarea, research method used, and government agency involved.    \\ \hline
Synthesis Procedure  & See the preceding point.  \\ \bottomrule          
\end{tabular}
\end{table}

\begin{table}[htbp!]
	\caption{Key features of the RR on government SE practices in regional venues (RR2).}
	\label{tab:2rr}
\begin{tabular}{p{1.1cm}p{6.7cm}}
\toprule
Research Question    & What evidence on SE government practices was published in regional venues during 2024?    \\ \hline
Time Frame           & Approximately 40 hours.    \\ \hline
Search Strategy      & Manual search of papers published in 2024 in the following venues: Latin American Informatics Conference (CLEI), Ibero-American Conference on Software Engineering (CIbSE), and the Brazilian Symposium on Software Engineering (SBES).       \\ \hline
Selection Procedure  & We considered studies conducted by, for, or in collaboration with governments (mentions of government, specific countries, public agencies, regulatory bodies, etc.), in English, Spanish, or Portuguese, with full text available. The third author performed the study selection, and the first author revised the outcomes.  \\ \hline
Extraction Procedure & The third author guided by the first author collected publication data (title, authors, etc.), countries of author affiliations, and classified items by topics of interest, for example SE subarea, research method used, and government agency involved.      \\ \hline
Synthesis Procedure  & See the preceding point. \\ \bottomrule                                  
\end{tabular}
\end{table}

\begin{table}[htbp!]
	\caption{Key features of the RR of government SE practices in public-sector digitalization publications (RR3).}
	\label{tab:3rr}
\begin{tabular}{p{1.1cm}p{6.7cm}}
\toprule
Research Question    & What evidence on government SE practices appears in government digitalization publications?     \\ \hline
Time Frame           & Approximately 30 hours.    \\ \hline
Search Strategy      & Manual search. We reviewed all 2024 issues of the EDOS newsletter (Effective Digitalization of the Public Sector). EDOS is a weekly digital newsletter that shares knowledge about successful public-sector digitalization with Norwegian government agencies. It is a research center led by scholars from Norway’s IT management field and funded by the Ministry of Digitalization and Public Governance.      \\ \hline
Selection Procedure  & We included studies conducted by, for, or in collaboration with governments (mentions of government, specific countries, public agencies, regulatory bodies, etc.), written in English, with full text available. EDOS published 53 issues in 2024. Each issue typically lists six studies, three as main items and three as bonus items. For this rapid review we selected only the main studies. The second author performed the study selection, and the first author revised the outcomes.     \\ \hline
Extraction Procedure & The second author extracted the following data: authors, title, year, abstract, and venue.     \\ \hline
Synthesis Procedure  & We categorized studies using: countries involved (both authors’ affiliations and the study context), topic area (SWEBOK classification), study type (Wieringa’s classification), and knowledge discipline. For the classification, the second author used ChatGPT and then manually reviewed the outputs. The first author independently classified a sample of papers. \\ \bottomrule
\end{tabular}
\end{table}

\begin{figure*}[htbp!]
\begin{center}
\includegraphics[width=16cm]{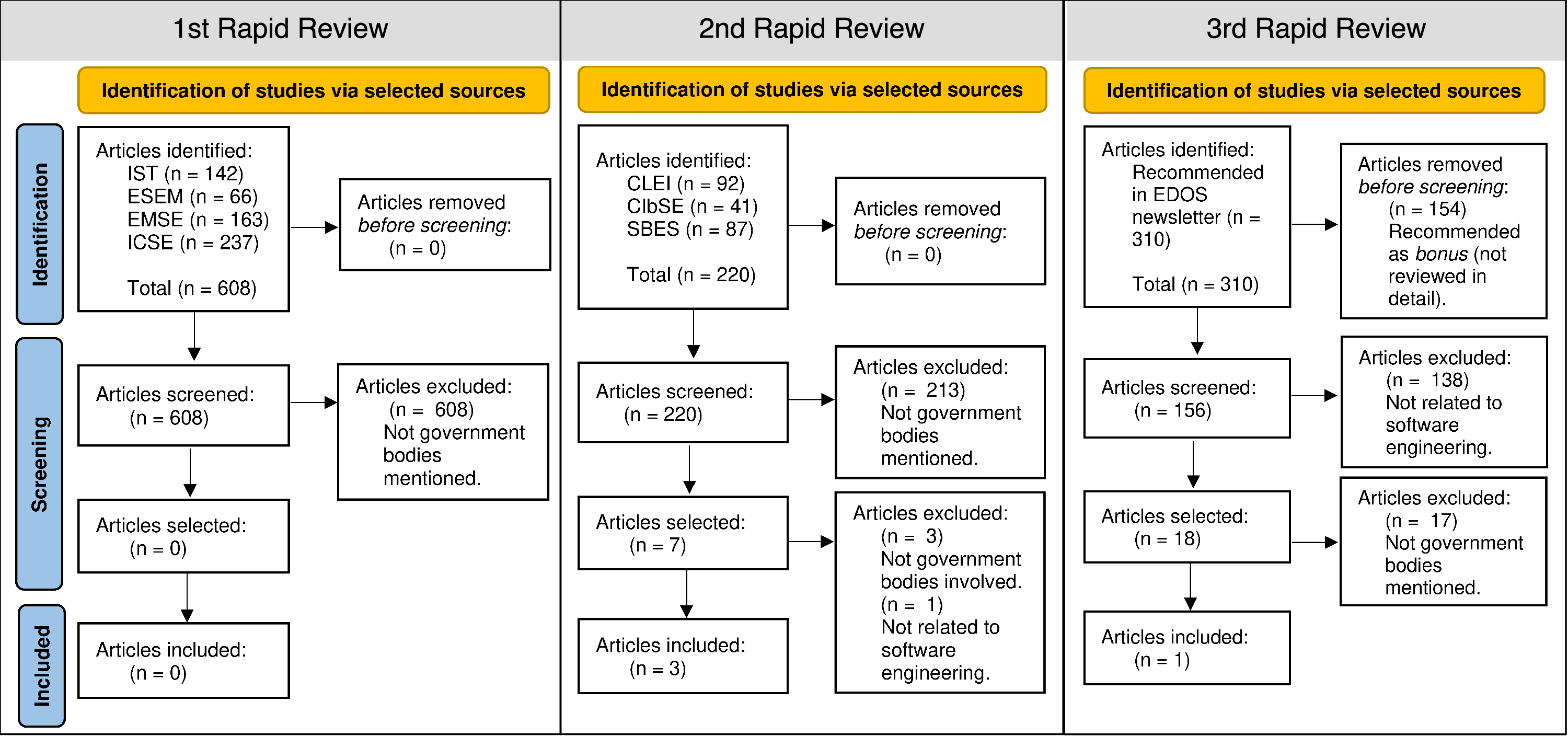}
\end{center}
\caption{PRISMA Flow diagrams of the conducted RRs.} \label{flows}
\end{figure*}

Figure \ref{flows} presents the PRISMA flow diagrams \cite{page2021} for the three RRs. In the RR1, which targeted top-tier SE venues, we screened 608 papers and found that none mentioned government bodies. In the RR2, we initially identified seven papers that mentioned governments, but the governments’ roles were unclear. After contacting the authors, we confirmed that only four papers involved some collaboration with the government bodies mentioned. One paper was excluded at a late stage because it reported operations-research practices rather than SE practices \cite{blasiak2024}. In RR3, we considered only the 156 publications reviewed in detail within the EDOS newsletter issues, excluding items listed solely for further reading (labeled as “bonus”). Because the publications covered heterogeneous knowledge areas, we first used ChatGPT to classify all items and then selected those labeled as SE in a second step. Prior work suggests that LLMs can be quite accurate for this type of simple, descriptive classification and qualitative synthesis \cite{pizard2025usellms}. In addition, after the second author performed the ChatGPT-assisted classification, the first author validated the results on a random sample. Beyond the need to contact authors in RR2, we made no other major deviations.

\section{Findings \& Discussion} \label{sec:results}

\begin{table*}[htbp!]
	\caption{Papers found on government software engineering practices published in 2024.}
	\label{tab:papers}
\begin{tabular}{lp{1cm}p{2.7cm}p{4.3cm}p{1.2cm}p{1.2cm}p{1.2cm}p{2.5cm}}
\toprule
Study       & Countries Involved & Government Bodies Involved & Study Purpose & SE Topic & Research Method   & Source         & Other comments \\ \midrule 
\cite{Heikel2024}  & Paraguay           & Ministry of Economy and Finance & Adapting and integrating a microservices architecture into Paraguay's state financial system (SIARE), replacing a monolithic client--server setup to improve scalability and reduce coupling. & Software Architecture        & Case Study        & CLEI (RR2)  & Authors from the Ministry's IT directorate. In Spanish, 8 pages. \\
\cite{Costa2024}   & Brazil             & Fortaleza City Hall (Municipal Government of Fortaleza, Ceará State, Brazil) & Experience report on the difficulties and solutions when specifying requirements for a public-sector Big Data platform. & Software Requirements        & Experience report & CIBSE (RR2) & Applied R\&D track, 4 pages. In Portuguese. \\
\cite{Batista2024} & Brazil             & Fortaleza City Hall (Municipal Government of Fortaleza, Ceará State, Brazil) & The Big Data Fortaleza platform, applying Big Data and ML to support strategic planning for sustainable urban development (initially early-childhood). & Software Architecture        & Experience report & CIBSE (RR2) & Applied R\&D track, 4 pages. \\
\cite{McInnes2024} & USA                & U.S. DOE (Office of Science and NNSA); national-level impact. & How the DOE-funded IDEAS project (Interoperable Design of Extreme-scale Application Software) improves scientific-software productivity and sustainability for extreme-scale computing by building communities and disseminating effective practices. & Software Engineering Process & Experience report & EDOS (RR3) \textit{Computing in Science \& Engineering} & 30 authors from universities and national laboratories. 13 pages. \\ \bottomrule            
\end{tabular}
\end{table*}

This section presents the results of our study, additional preliminary investigations that complement our main work, and some reflections on its implications.

\paragraph{Summary of Findings} Across the three RRs, we screened 984 papers and identified only four studies conducted by, for, or in collaboration with government bodies. Given the brevity of the included papers (mostly short experience reports), the practitioner-facing detail they offer is limited; we summarize their key features in Table \ref{tab:papers}.

Three of the papers are described by their authors as experience reports, and one as a case study, although recent literature suggests it could also be classified as an experience report \cite{wohlin2021b}. The topics covered include: software architecture, software requirements, and software engineering process. Two of the papers are short, published as extended abstracts. These two were presented at CIBSE 2024, held in Curitiba, Brazil, and both report collaborations with a nearby municipality. Of the four papers, two are written in English, one in Spanish, and one in Portuguese. We did not formally assess quality, as this generally falls outside the scope of RRs. Looking informally at the papers, and perhaps because they are short and mostly experience reports (without a known empirical method as a basis), they appear not to qualify as high-quality studies.

Although the sample is very small, these features may hint at factors that could encourage government-related publications: not requiring sophisticated or highly rigorous research methods; venues located geographically close to the government agencies involved; allowing authors to publish in their native language; and shorter publication formats. However, such short formats provide limited support for readers who wish to replicate the reported results in their own contexts, due to the lack of details.

Among the SE and computing venues examined (refer to RR1 and RR2), only an ESEM track explicitly calls for government-related submissions: the Industry, Government and Community Track (IGC)\footnote{https://conf.researchr.org/track/esem-2024/esem-2024-industry-government-community?}. ICSE has the Software Engineering in Practice (SEIP) track, but its description does not refer to government; in particular, it states: \textit{``SEIP will include participants and speakers from both industrial and academic sectors."} CIBSE has an industry talks track, but it does not publish papers or abstracts. While the thematic scope of CLEI is much broader than software engineering, it does include a track on Systems in Practice, focusing on Computer Systems Deployed in Productive or Social Environments. This track covers a variety of topics, including: Human-Computer Interaction, High-Performance Computing, Bioinformatics, Chemoinformatics, and Applications in Industry, Government, and Society.

The following reflections draw on the authors' experience collaborating with government agencies rather than on the data collected in the RRs, and should be read as hypotheses to be examined in future work. Our experience collaborating with government agencies suggests that the main motivation for these initiatives is usually to improve the agencies’ practices, and that neither research nor the publication of experience reports is prioritized by agency staff or authorities. Much of the research we have seen published jointly with government agencies focuses on issues that have a direct impact on citizens, which generally does not apply to topics in our discipline. In addition, whenever we have explicitly included research activities in our collaboration proposals, it has been necessary to clearly explain our intentions and agree on the terms of the work (especially regarding the publication of results). These confidentiality concerns add to the difficulty we perceive as researchers during the review process when trying to publish field studies compared to other types of research.

\paragraph{Further Investigations} The following two activities are exploratory and complementary to our main RQ; they do not answer it directly but motivate future directions. 

In a previous study on evidence-based practice adoption \cite{pizard2023}, participants not only identified the lack of evidence as a potential barrier but also reported that they currently rely heavily on grey literature for decision-making. We decided to broaden our investigation to better understand and improve the production and use of this type of material by government agencies. To begin exploring this, we carried out two activities. First, we reviewed the websites of some regional government digitalization agencies to identify the publications they produce, and, second, we also trialed a checklist to assess grey literature for decision-making support.

\textbf{Grey literature published by government agencies.} We reviewed the online publications of government agencies, AGESIC in Uruguay and similar agencies in Argentina, Paraguay, and Chile, with the goal of identifying what types of documents they published during 2024\footnote{Sources consulted by country: Uruguay: AGESIC, https://www.gub.uy/agencia-gobierno-electronico-sociedad-informacion-conocimiento/comunicacion/publicaciones. Argentina: Agency for Access to Public Information (AAIP), https://www.argentina.gob.ar/aaip/documentos-de-acceso-la-informacion/ and Secretariat for Public Innovation (SIP), https://www.argentina.gob.ar/jefatura/innovacion-ciencia-y-tecnologia/. Paraguay: Ministry of Information and Communication Technologies (MITIC), https://mitic.gov.py/gobierno-electronico/. Chile: Digital Government Division (DGD), https://datos.gob.cl/organization/division-de-gobierno-digital/.}. For this supplementary review of agency websites only (not for the main rapid reviews, which did include Brazilian venues and Portuguese-language papers) we did not cover Brazil's agency, due to resource constraints.

\begin{figure}[htbp!]
\begin{center}
\includegraphics[width=7cm]{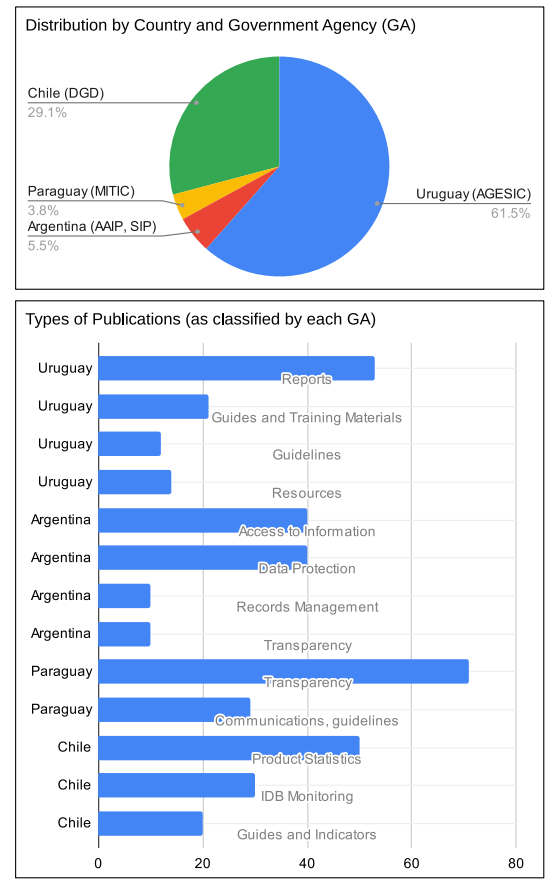}
\end{center}
\caption{Publications of Regional Digital Government Agencies during 2024.} \label{greymaterials}
\end{figure}

We identified publications produced in 2024 by these digital government bodies; Figure~\ref{greymaterials} summarizes the results by document type. Output varied widely: Paraguay (MITIC, 7 documents) and Argentina (AAIP, 10) published very little (and the Argentine SIP website showed no 2024 items), whereas Chile (DGD, 53) and especially Uruguay (AGESIC, 112) published substantially more. The document mix also differed by country: reports and guides dominate in Uruguay, access-to-information and data protection in Argentina, and product statistics and IDB monitoring in Chile (see Fig.~\ref{greymaterials} for the full breakdown). Whether state policies (such as science funding (or defunding) or initiatives promoting management transparency) explain these differences is an interesting question, but a causal analysis is beyond the scope of this exploratory study and is left as future work.

\textbf{Quality assessment of SE grey literature.} Given the limited academic literature we identified, and the finding by \cite{pizard2023} that government agencies' staff use grey literature, we decided to explore how to support them in considering this type of information appropriately. In particular, we started investigating whether the quality assessment checklists for grey literature proposed by \cite{garousi2019} could help practitioners.

\begin{table}[t]
\caption{Checklist to assess grey literature adapted from \cite{garousi2019}.}
\label{tab:grey-literature-checklist}
\centering
{
\begin{tabular}{p{0.95\columnwidth}}
\hline
\textbf{Producer authority} \\
\hline
- Is the publishing organization reputable? E.g., the Software Engineering Institute (SEI) \\
- Is the author affiliated with a prestigious organization? \\
- Has the author published other work in the field? \\
- Does the author have experience in the area? (e.g., job title such as principal software engineer / technical lead) \\
\hline
\textbf{Methodology} \\
\hline
- Does the source have a clearly stated objective? \\
- Does the source have a clearly specified methodology? \\
- Is the source supported by references to authoritative (reliable) and contemporary sources? \\
- Are the limitations of the work clearly stated? \\
- Does the work address a specific question? \\
- Does the work refer to a particular population or case? \\
\hline
\textbf{Objectivity} \\
\hline
- Does the work appear balanced in its presentation? \\
- Are the claims in the source as objective as possible, or are they subjective opinions? \\
- Is there any personal interest? E.g., a tool comparison conducted by authors who work for a specific tool vendor \\
- Are the conclusions supported by the data? \\
\hline
\textbf{Date} \\
\hline
- Does the document have a clearly indicated date? \\
\hline
\textbf{Position relative to related sources} \\
\hline
- Have key grey literature sources or related formal sources been linked or discussed? \\
\hline
\textbf{Novelty} \\
\hline
- Does it enrich the research or add something unique? \\
- Does it strengthen or refute a current position? \\
\hline
\textbf{Impact} \\
\hline
- Normalize all the following impact metrics into a single aggregated impact metric (when data are available): number of citations, number of inbound links, number of social media shares (so-called ``altmetrics''), number of published comments on specific online entries such as a blog post or a video, number of page views or document views. \\
\hline
\textbf{Medium type} \\
\hline
- What type of literature do you think this is? \\
-- 1st tier grey literature (score = 1): High outlet control / High credibility: books, magazines, theses, government reports, white papers. \\
-- 2nd tier grey literature (score = 0.5): Moderate outlet control / Moderate credibility: annual reports, news articles, presentations, videos, Q\&A sites (such as StackOverflow), wiki articles. \\
-- 3rd tier grey literature (score = 0): Low outlet control / Low credibility: blogs, emails, tweets. \\
\hline
\end{tabular}
}
\end{table}

As a preliminary investigation, we invited the 22 students from an EBSE and SRs course to complete a questionnaire based on the checklist in \cite{garousi2019} (see Table~\ref{tab:grey-literature-checklist}). Students could use the questionnaire to assess information sources they were considering for decision-making in their professional practice. We also asked them which problem they were addressing, to provide additional details about the source, and to share feedback on the checklist, including whether they found it appropriate and useful.

Three respondents participated. Two addressed technical problems (implementing an AWS Lambda connection to a database hosted on AWS, and programmatically managing Google Ads keywords and ad campaigns via the Java API), while the third focused on analyzing user behavior on a government agency website to improve usability and intuitiveness. Overall, respondents considered the checklist useful. However, two of the three respondents (those working on the more technical problems) stated that they would not use the checklist in the future because it is too time-consuming, and suggested merging or synthesizing questions. One of them commented: \textit{``[...] it would take me quite a lot of time to implement this evaluation for every question I have.''}

Although this is a preliminary inquiry with a very small sample, the results help us refine our direction. Even if the checklist still needs improvement, it appears promising for supporting decision making when using grey literature. However, we need to better understand and clearly define when we should recommend using this type of assessment (e.g., highly technical or coding-related problems may not be well suited to this approach). 
To tackle this issue and reduce the time needed to utilize the checklist, it is particularly beneficial to explore the use of Large Language Models (LLMs). For example, we could create one or a set of prompts that integrates the checklist and enables semi-automated or automated analysis of grey literature.

\paragraph{Implications for Practice \& Research} The concrete guidance we can offer SE practitioners working in government agencies is limited by the exploratory nature and small evidence base of this study. We did not find venues within the SE community that clearly included studies of government practices. However, the three papers on GAs' SE practices that we identified in SE venues were all published in regional conferences. This suggests that regional venues may be more receptive to publications of this type. We plan to continue investigating this topic. In the meantime, we recommend that practitioners use and refine methods to evaluate the quality of the grey literature they actually rely on. We do not propose grey literature as a substitute for peer-reviewed evidence; rather, since practitioners already rely on it \cite{pizard2023}, we explore how to help them appraise it more rigorously. The checklist presented in Table~\ref{tab:grey-literature-checklist}, based on the work of \cite{garousi2019}, may be useful for this purpose.

For our research colleagues, we want to raise awareness of this situation. Our preliminary findings suggest that academic evidence on government SE practices has low visibility in the venues we examined. Efforts could be made to address this gap. For example, by encouraging publications in this area through journal special issues on GAs' SE practices, or by fostering collaboration between researchers and communities such as the UK cross-government SE community \cite{ukgov2025}.

\section{Threats to Validity}\label{sec:threats}

The main research of this study (i.e., the planning and execution of the three RRs) faces several validity threats that we analyze and present according to Ampatzoglou's proposal for secondary studies \cite{Ampatzoglou2019}.

\textbf{Study Selection Validity.} This category involves threats that can be identified in the first two phases of secondary studies planning (i.e., search process and study filtering phase). In this regard, the sources we surveyed were selected according to our criteria which, although explained previously and intended to be objective, may contain biases. Manual screening over many publications introduces a risk of reviewer fatigue and missed items; to partially mitigate this, screening was distributed across reviewers and a sample of decisions was validated by the first author.

\textbf{Data Validity.} This category encompasses threats identified during the final two phases of secondary studies (specifically, data extraction and analysis) that can jeopardize the validity of both the extracted dataset and its subsequent analysis. Data extraction was primarily conducted by the second and the third authors, with support and validation from the first author. In the first review report (RR1), the second author utilized ChatGPT to classify the preselected papers into predefined and known categories (i.e., SE topic and research method). This type of classification falls under descriptive synthesis, which is currently recommended for Large Language Models (LLMs) support at this stage of SRs and RRs (refer to \cite{pizard2025usellms}). Nevertheless, to mitigate the risks of errors or hallucinations, the first author reviewed a random sample of the classifications assisted by the model.

\textbf{Research Validity.} Threats that can be identified in all four phases and concern the overall research design are classified into this category. The main reviewers had limited experience conducting RRs. The second author had previously taken an EBSE and SRs course (the one reported in \cite{pizard2021}), in which he had participated in conducting an SR, whereas the third author was trained in EBSE and SRs prior to this research. Although they had been previously trained and worked closely with the first author, for both of them these were their first RRs. To mitigate possible deviations from RR methodology, we followed RR guidelines, including developing a protocol before conducting the review. Additionally, the first author played both the requester and lead roles in the RRs. To mitigate potential impacts on methodological rigor, we documented all decisions made and any deviations from the protocols, and we involved the fourth author, who reviewed the conducted RRs and helped discuss their results. Consistent with rapid-review practice, we did not perform a formal quality assessment, screening relied predominantly on a single reviewer with verification by the first author rather than independent double screening, and we did not compute inter-rater reliability. These choices favor speed over the rigor of a full systematic review and should be considered when interpreting the results. An additional threat concerns the use of a single publication year and a limited set of sources. Consequently, the findings should be interpreted as an exploratory assessment rather than a definitive characterization of the entire evidence base on government software-engineering practices. The results are intended to indicate whether evidence is readily visible in selected SE contemporary sources, not to estimate the total volume of relevant studies published worldwide.

\section{Final remarks} \label{sec:final}

In this work, we conducted an initial exploration of the academic evidence on SE practices conducted for, by, or in collaboration with government agencies or bodies. Specifically, we manually reviewed leading venues in empirical SE, regional venues in SE and computing, and a newsletter that reviews publications on government digitalization. We found very few publications explicitly reporting software-engineering practices conducted by, for, or in collaboration with government agencies (4 out of nearly one thousand screened publications). This finding suggests that such evidence is either scarce, dispersed across other publication venues, or insufficiently visible within the sources examined in this study.

Given the exploratory nature of this study, several opportunities exist to expand and refine the investigation. As future work, it would be interesting to: (1) manually sample venues focused on government digitalization as well\footnote{The venues that we consider particularly relevant are the following. For journals: Government Information Quarterly (GIQ), Transforming Government: People, Process and Policy (TGPPP), Information Polity, and Digital Government: Research and Practice (DGOV, ACM). For conferences: IFIP EGOV-CeDEM-ePart, dg.o – International Conference on Digital Government Research (Digital Government Society), and the HICSS – Digital Government track.}; (2) use the experience gained in this research to design and conduct a broader search (e.g., the last five or ten years), likely automated, to identify publications on government SE practices; (3) investigate which information sources government bodies use for decision-making, particularly when developing regulations, standards, and policies, and (4) develop guidelines and training materials that enable government agencies themselves to generate their own evidence with a minimum level of rigor appropriate for its later use in decision making.

\bibliographystyle{IEEEtran}
\bibliography{main}

\end{document}